\documentclass[twocolumn]{aastex701}

\newcommand{\re}{R_{\oplus}}
\newcommand{\ms}{M_{\odot}}
\newcommand{\sbol}{S_{\oplus}}
\newcommand{\dx}{\Delta_{\mathrm{M-FGK}}}
\newcommand{\phix}{\Phi_{\mathrm{XUV}}}
\graphicspath{{figures/}{./}}
\ifdefined\showrevision
  \newcommand{\rev}[1]{\textbf{#1}}
\else
  \newcommand{\rev}[1]{#1}
\fi

\shorttitle{Does Cumulative XUV Unify the Neptunian Desert?}
\shortauthors{Jiang}
\accepted[Accepted for publication in ]{The Astronomical Journal}

\begin{document}

\title{Does Cumulative XUV Unify the Neptunian Desert?}

\correspondingauthor{Qunfeng Jiang}
\author[0009-0008-0600-3743]{Qunfeng Jiang}
\affiliation{Independent Researcher, Shanghai, China}
\email[show]{qfjiang01@gmail.com}

\begin{abstract}
\rev{The Neptunian desert is the scarcity of roughly $2$--$10\,\re$ planets on orbits of a few days.  Atmospheric escape driven by stellar X-rays and extreme-ultraviolet (XUV) radiation may shape its lower boundary by stripping hydrogen--helium envelopes.  Because M dwarfs and F-, G-, and K-type stars have different activity histories, present bolometric irradiation may not trace the cumulative high-energy dose that drives escape.  We ask whether cumulative XUV yields a common boundary across spectral type and whether the convergence survives star-to-star variation in age and rotation.  Using 1850 Kepler and TESS planets, Model for Rotation of Stars (MORS) activity tracks, and a global boundary model, we find that the M-dwarf--FGK offset shrinks from $-1.065$ dex in bolometric flux to $-0.369$ dex for a common 5 Gyr median-rotation XUV history.  Assigning plausible ages and initial rotations to individual hosts preserves the smaller XUV offset in every trial, with a median reduction of 69.1\%.  A common-period null yields comparable alignment.  Cumulative XUV therefore substantially reduces the host-type dependence, but the present data do not establish a universal XUV threshold independent of orbital period.}
\end{abstract}


\section{Introduction}\label{sec:intro}

The discovery of thousands of close-in planets revealed a strongly structured period--radius distribution.  One prominent feature is the Neptunian desert, a scarcity of roughly $2$--$10\,\re$ planets at periods of a few days or less.  Early surveys identified the deficit \citep{Szabo2011,Beauge2013}, and \citet{Mazeh2016} measured its upper and lower boundaries.  Short-period planets of these sizes are readily detected, making transit sensitivity an unlikely origin of the feature.  Population maps also show a ``ridge'' near 3--6 days and a less depleted ``savanna'' at longer periods \citep{CastroGonzalez2024}.

Atmospheric loss offers a physical origin for the desert's lower boundary \citep{OwenLai2018,Hallatt2022}.  X-ray and extreme-ultraviolet (XUV) photons heat a hydrogen--helium atmosphere and can drive a hydrodynamic outflow.  Planets with modest cores and envelopes may be stripped to smaller remnants, while deeper gravitational potentials favor survival \citep{OwenWu2013,LopezFortney2013,Owen2019}.  Related calculations reproduce broad features of the small-planet radius distribution \citep{Fulton2017}.  Tidal survival, high-eccentricity migration, and runaway mass loss are more relevant to the upper desert boundary \citep{Matsakos2016,OwenLai2018,Ionov2018,Vissapragada2022,Thorngren2023}; high-eccentricity tidal migration can also generate the adjacent ridge \citep{CastroGonzalez2026}.  These mechanisms need not leave the same dependence on host-star properties.

Present bolometric flux records today's stellar luminosity and orbital separation, whereas atmospheric escape responds to high-energy irradiation accumulated over time.  Magnetic activity depends on rotation and decays as stars spin down; both the saturation lifetime and $L_{\rm XUV}/L_{\rm bol}$ vary with stellar mass \citep{Jackson2012,Wright2011,Shkolnik2014,Johnstone2021}.  \rev{The Sun in Time and HAbitable Zones and M dwarf Activity across Time (HAZMAT) programs show that high-energy emission declines with age, while low-mass stars retain stronger and more dispersed activity for longer \citep{Ribas2005,Schneider2018,RicheyYowell2019,Peacock2020,RicheyYowell2023}.  These observations motivate an integrated XUV coordinate for comparisons across spectral type.}  M dwarfs are bolometrically faint and can remain magnetically active far longer than solar-type stars, especially near and below the fully convective boundary \citep{Pass2025}.  The integral of $L_{\rm XUV}/(4\pi a^2)$ is therefore a physically motivated measure of the irradiation available to drive escape.  Direct extreme-ultraviolet (EUV) histories require reconstruction because interstellar absorption blocks most stellar EUV emission \citep{SanzForcada2011,France2016}.

Measurements of the desert across stellar populations already show strong coordinate dependence \citep{SzaboKalman2019,Szabo2023}.  Lifetime X-ray irradiation yields a more coherent $1.8$--$4\,\re$ sub-Neptune boundary across spectral type than bolometric irradiation \citep{McDonald2019}.  In a larger Kepler--Transiting Exoplanet Survey Satellite (TESS) sample, \citet{Osborn2026} found opening periods of $2.2\pm1.0$ days around M dwarfs and $3.3\pm1.4$ days around F-, G-, and K-type (FGK) stars, together with bolometric-flux scales of approximately $58$ and $831\,\sbol$.  These studies establish the motivation for an XUV coordinate, but coordinate alignment alone cannot identify atmospheric erosion.  At fixed period, stellar mass sets the orbital separation and also enters the modeled activity history.  A nearly common period edge can therefore map into a compact cumulative-XUV edge even when the planet population supplies no additional response to XUV.

We calculate standardized cumulative XUV exposures for the Kepler--TESS population, fit the M-dwarf and FGK desert boundaries with one global model, and test the result against a common-period mapping null.  This design measures both the empirical convergence and the part expected from stellar-mass-dependent coordinate transformation.  Section~\ref{sec:methods} presents the sample, stellar histories, boundary estimator, and null construction.  Section~\ref{sec:results} gives the offsets and sensitivity tests.  Section~\ref{sec:discussion} connects those measurements to atmospheric loss and identifies observations that can break the period--XUV degeneracy.  \rev{Section~\ref{sec:conclusions} summarizes the conclusions.}

\section{Data and Methods}\label{sec:methods}

\subsection{Planet sample}\label{subsec:sample}

We use the public planet table of \citet{Osborn2026}, assembled from the Kepler \citep{Borucki2010} and TESS \citep{Ricker2015} surveys.  The table contains 2786 entries.  We required a reported transit signal-to-noise ratio (S/N) of at least 7.1, removed 44 cross-survey duplicates by matching periods and retaining the Kepler measurement, and selected $P<30$ days and $1<R_p/\re<25$.  We restricted stellar masses to $0.1<M_*/\ms<1.2$, the interval covered by the activity calculations in Section~\ref{subsec:xuv}.  The resulting sample contains 1850 planets around 1462 hosts: 1430 Kepler planets (55 M-dwarf and 1375 FGK) and 420 TESS planets (66 M-dwarf and 354 FGK).  We classify hosts with $T_{\rm eff}<3900$ K as M dwarfs, consistent with \citet{Osborn2026}; stellar masses span $0.139$--$1.198\,\ms$.

The global boundary fit uses the individual planets with $2.24<R_p/\re<7.48$.  This interval closely matches bins 3--5 of the eight-bin reference grid and brackets the observed desert opening \citep{Osborn2026}.  It contains 677 planets around 581 hosts, including 36 M-dwarf planets (16 Kepler and 20 TESS) and 641 FGK planets (557 Kepler and 84 TESS).  Fits over $2.24$--$5\,\re$ and $2$--$8\,\re$ test the radius selection.

Each object has a statistical weight
\begin{equation}
 w_i = \left(p_{{\rm tr},i}\,p_{{\rm det},i}\right)^{-1},
 \label{eq:weight}
\end{equation}
where $p_{\rm tr}$ is the geometric transit probability and $p_{\rm det}$ approximates signal recovery.  The inverse factors allow a less probable detection to represent a larger number of underlying systems.  Kepler's recovery curve is calibrated by transit injection tests \citep{Christiansen2020}.  The heterogeneous TESS list lacks one corresponding end-to-end selection function, and its weights use the same efficiency curve as an approximation \citep{Osborn2026}.  We report empirical boundaries of the weighted detections.  Survey-separated and unit-weight fits show how this approximation affects the result.

\subsection{Cumulative XUV exposure}\label{subsec:xuv}

We calculated stellar X-ray and extreme-ultraviolet (EUV) histories with the Model for Rotation of Stars (MORS; \citealt{Johnstone2021}), using the stellar-structure tracks of \citet{Spada2013}.  MORS evolves rotation and magnetic activity for $0.1$--$1.2\,\ms$ stars and returns $L_{\rm X}(t)$ and $L_{\rm EUV}(t)$.  The tracks include saturated high-energy emission, spin-down, and the early dispersion produced by stellar rotation \citep{Tu2015,Johnstone2021}.  We evaluated 51 uniformly spaced mass nodes and interpolated in $\log E_{\rm XUV}$.

For stellar mass $M_*$, integration endpoint $t_f$, and rotation track $r$, we defined
\begin{eqnarray}
 E_{\rm XUV}(M_*,r,t_f) &= \int_{10\,{\rm Myr}}^{t_f}
       \left[L_{\rm X}(t)+L_{\rm EUV}(t)\right]dt, \label{eq:energy}\\
 \phix &= \frac{E_{\rm XUV}}{4\pi a^2}. \label{eq:fluence}
\end{eqnarray}
We begin the integral at 10 Myr, when most primordial disks have dispersed and the close-in orbit is less coupled to a disk-embedded phase.  Starts at 5 and 20 Myr alter the inferred compression by only 0.2 percentage points (Appendix~\ref{app:support}).  Direct MORS calculations between mass nodes agree with the interpolation within 0.009 dex.

The rotation grid uses the 5th, 50th, and 95th percentile tracks.  These correspond to the slow tail, median, and fast tail of the calibrated rotation distribution and enclose its central 90\%; \citet{Johnstone2021} anchored the tracks to open-cluster rotation percentiles at 150 Myr.  Slow rotators leave X-ray saturation earlier, while fast rotators retain saturated emission longer.  The three tracks sample both tails and the population center, which is sufficient for testing whether the direction of boundary convergence depends on early rotation.  We integrate each track to 1, 3, and 5 Gyr.  These endpoints represent a young field population, an intermediate-age population, and a mature population near the solar age.  They also span the interval over which the model changes from widespread mass-dependent saturation to near-universal unsaturated emission above the late-M regime \citep{Johnstone2021}.  The baseline adopts the 50th-percentile track at 5 Gyr; the other eight combinations define the activity-history sensitivity grid.

\rev{We call $\phix$ standardized when all hosts share the same integration endpoint and initial-rotation percentile, while MORS retains the mass dependence of stellar activity.  The 5 Gyr, 50th-percentile track defines the baseline comparison.  Results labeled standardized use this common history; the host-level experiment below relaxes that assumption.}

\rev{To test the effect of heterogeneous stellar histories, we perform 2000 host-level Monte Carlo realizations.  Each host receives one age and one MORS initial-rotation percentile, and all planets in a multiplanet system share that draw.  The initial-rotation percentile is sampled uniformly between the 5th and 95th percentiles.  We do not draw a present-day rotation period separately: MORS evolves the rotation and activity history from the sampled initial state, stellar mass, and age.  Integrated XUV energies are interpolated over seven initial-rotation percentiles and ages of 1, 3, and 5 Gyr.}

\rev{The primary age prior draws FGK hosts from the 538 of 549 fitted-sample FGK hosts with valid catalog ages \citep{Berger2023}.  The M-dwarf prior uses the 25 of 32 fitted-sample M-dwarf hosts with literature or gyrochronology ages in the compilation described in Appendix~\ref{app:support}.  The reported ages remain in the empirical pools; whenever a sampled age lies outside the 1--5 Gyr MORS interpolation interval, it is represented by the nearest endpoint.  A sensitivity calculation draws both host classes from the pooled age sample.  Each realization recomputes $\phix$ and refits Equation~\ref{eq:boundary}; measured planet parameters and statistical weights remain fixed.}

\subsection{Boundary inference and null model}\label{subsec:boundary}

As a nonparametric visualization, we divide planet radius into eight logarithmic bins from 1 to $25\,\re$ and estimate an edge separately in each bin.  The binning and percentile conventions follow \citet{Osborn2026}.  For each host class, the short-period edge is the weighted 2.3rd percentile of $P$; the high-irradiation edge is the weighted 97.7th percentile of bolometric flux or $\phix$.  Figure~\ref{fig:spaces} displays bins 3--5 ($2.236$--$7.477\,\re$), which differ negligibly from the $2.24$--$7.48\,\re$ global-fit interval.  The small M-dwarf counts in individual bins motivate a single boundary model across this interval.

Our primary estimator fits $2.24<R_p/\re<7.48$.  For coordinate $x$, define $y=\log_{10}x$ for flux and XUV, and $y=-\log_{10}P$ for period so that the relevant edge is always an upper quantile.  The symbol $q$ denotes the conditional quantile fitted by the model: $q=0.977$ traces the upper 97.7th percentile of $y$, equivalent to a one-sided $2\sigma$ edge for a Gaussian distribution.  We fit
\begin{equation}
 y_i=f(\log R_{p,i})+\gamma I_{{\rm TESS},i}
       +\dx I_{{\rm M},i}+\epsilon_i,                     \label{eq:boundary}
\end{equation}
where $f$ is a continuous linear hinge spline centered at $4\,\re$ with knots at 3 and $5\,\re$.  M and FGK hosts share its shape; $\dx$ is their fitted offset at fixed radius and survey.  Quantile regression minimizes weighted pinball loss.  Robustness fits use $q=0.95$, 0.97, 0.977, and 0.99; no, one, or three alternative knots; and a model with an additional $I_{\rm M}\log R_p$ term that permits different host-class slopes.

Uncertainties come from 2000 stratified host-bootstrap realizations.  Within each survey--host-class cell we resample hosts, retaining multiplanet systems intact.  A stellar mass and luminosity are drawn once per host and shared by all its planets.  Planet period and radius are then perturbed, $a$ is recomputed from the drawn $M_*$ and $P$, and both bolometric flux and XUV follow from that same $a$.  Transit and recovery weights are also updated.  The shared host-level draws preserve the physical dependence among $M_*$, $a$, and irradiation.  We quantify convergence as
\begin{equation}
 C=1-\frac{|\dx({\rm XUV})|}{{|\dx({\rm bol})|}},          \label{eq:compression}
\end{equation}
so $C=1$ denotes complete removal of the bolometric host offset, $C=0$ no improvement, and $C<0$ a larger separation.  \rev{Thus, $100C$ gives the percentage reduction in the logarithmic M-dwarf--FGK boundary separation when bolometric flux is replaced by cumulative XUV.}  Because $C$ becomes unstable when a simulated bolometric offset approaches zero, we also use the absolute reduction
\begin{equation}
 D=|\dx({\rm bol})|-|\dx({\rm XUV})|,                     \label{eq:reduction}
\end{equation}
measured in dex.

The period-mapping null begins with Equation~\ref{eq:boundary} fitted in period with $\dx=0$, producing a common surface $\widehat{y}_P(R_p,{\rm survey})$.  We define the residual in the fitted period coordinate as $r_i=y_{P,i}-\widehat{y}_{P,i}$.  For each realization $b$, the residuals are randomly permuted within survey and radius bin; $r_i^{(b)}$ denotes the residual assigned to planet $i$.  The simulated coordinate and period are
\begin{equation}
 y_{P,i}^{(b)}=\widehat{y}_{P,i}+r_i^{(b)}, \qquad
 P_i^{(b)}=10^{-y_{P,i}^{(b)}}.                            \label{eq:nullperiod}
\end{equation}
For every realization, $P_i^{(b)}$ and the observed $M_*$ determine a new $a_i^{(b)}$.  We then recompute bolometric flux, XUV fluence, and geometric transit probability.  At fixed observing baseline and per-transit precision, the counterfactual signal-to-noise ratio scales as ${\rm S/N}^{(b)}={\rm S/N}\sqrt{P/P^{(b)}}$; the Kepler Data Release 25 (DR25) gamma recovery curve supplies a new detection weight.  Both $\dx({\rm bol})$ and $\dx({\rm XUV})$ are refitted within each realization before calculating $C$ and $D$.  We use 2000 weighted realizations, 2000 unweighted realizations, and 2000 weighted realizations after retaining one seeded-random planet per host.  The construction retains the observed stellar population, Keplerian geometry, selection weights, and the MORS mass scaling while imposing a common period edge; atmospheric evolution does not enter the calculation.

\section{Results}\label{sec:results}

\subsection{Apparent XUV convergence}\label{subsec:convergence}

\begin{figure*}[t!]
\centering
\includegraphics[width=\textwidth]{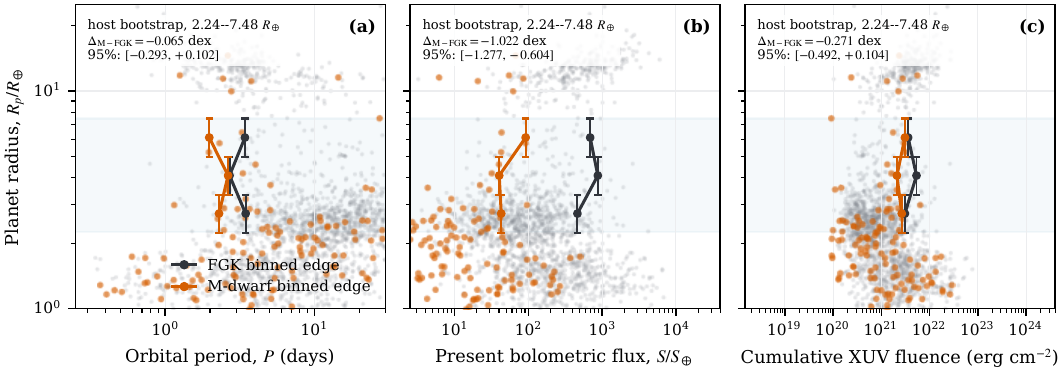}
\caption{The detected planet sample in orbital period, present bolometric flux, and standardized cumulative XUV fluence.  Gray points denote FGK hosts and orange points M-dwarf hosts.  \rev{The three connected symbols for each host class are weighted edge estimates in the $2.236$--$3.344$, $3.344$--$5.000$, and $5.000$--$7.477\,\re$ bins of the eight-bin reference grid.  Vertical bars show these radius-bin limits, not measurement uncertainties.}  The shaded band marks the $2.24$--$7.48\,\re$ global-fit interval.  \rev{Each inset reports the host-bootstrap median global M-minus-FGK offset and its central 95\% interval; Table~\ref{tab:global} also lists the point estimates.}}
\label{fig:spaces}
\end{figure*}

The M-dwarf and FGK boundaries nearly coincide in period.  Present bolometric flux separates them by about a decade, with the M-dwarf edge at lower irradiation.  Standardized cumulative XUV closes most of this gap (Figure~\ref{fig:spaces}).

Within $2.24$--$7.48\,\re$, the global model gives point offsets of $-0.013$, $-1.065$, and $-0.369$ dex in period, bolometric flux, and XUV, respectively.  \rev{Figure~\ref{fig:spaces} instead reports the corresponding host-bootstrap medians and central 95\% intervals.  The point compression is 65.4\%, meaning that cumulative XUV removes 65.4\% of the logarithmic M-dwarf--FGK boundary difference measured in present bolometric flux.  The host-bootstrap median of 73.4\% is the middle compression across 2000 samples formed by resampling host systems; it describes sensitivity to the observed host population.}  It differs from the point estimate because $C$ is a nonlinear ratio of two fitted offsets.  Period has the smallest host dependence, XUV is intermediate, and bolometric flux has the largest separation.

\begin{deluxetable*}{lcc}
\tablecaption{Host dependence of the lower-desert boundary\label{tab:global}}
\tablehead{
\colhead{Quantity} & \colhead{Point estimate} & \colhead{Bootstrap median [95\% interval]}
}
\startdata
$\dx(P)$ (dex) & $-0.013$ & $-0.065\;[-0.293,+0.102]$ \\
$\dx({\rm bol})$ (dex) & $-1.065$ & $-1.022\;[-1.277,-0.604]$ \\
$\dx({\rm XUV})$ (dex) & $-0.369$ & $-0.271\;[-0.492,+0.104]$ \\
$C$ (\%) & $65.4$ & $73.4\;[53.0,99.2]$ \\
$D$ (dex) & $0.696$ & $0.753\;[0.434,1.030]$ \\
\enddata
\tablecomments{The model uses 677 planets with $2.24<R_p/\re<7.48$.  Intervals come from 2000 host-bootstrap realizations with coherent stellar-parameter and orbital error propagation.}
\end{deluxetable*}

\subsection{Robustness of the convergence}\label{subsec:robust}

No individual host determines the XUV convergence.  Repeating the fit after removing each of the 581 hosts gives compression between 63.3\% and 94.2\%.  The result is also present in the denser $2.24$--$5\,\re$ interval ($C=69.1\%$), in the wider $2$--$8\,\re$ interval ($C=80.8\%$), and with equal weight assigned to every detected planet ($C=80.3\%$).

Kepler and TESS independently show the same direction, with $C=69.2\%$ and 93.2\%, respectively.  Alternative quantiles, boundary curvatures, and an M-dwarf-specific slope give compression from 59.3\% to 94.8\%.  \rev{The nine uniform stellar histories span 50.9\%--88.0\% (Figure~\ref{fig:histories}), and all nine give $C>0$.  The assumed rotation and integration age therefore affect the amplitude while preserving the convergence.}  Moving the integration start between 5 and 20 Myr changes $C$ by 0.2 percentage points.

\begin{figure}[t!]
\centering
\includegraphics[width=\columnwidth]{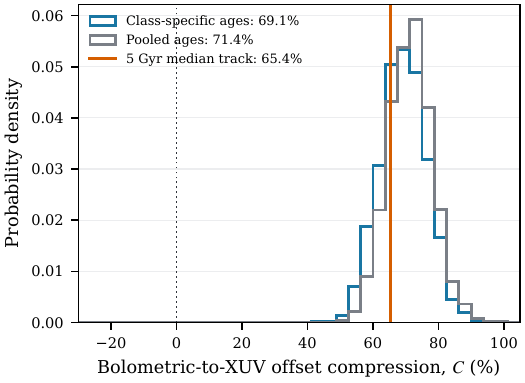}
\caption{\rev{Compression from host-level random stellar histories.  Each distribution contains 2000 realizations in which every host receives one age and one MORS initial-rotation percentile; multiplanet systems share the host draw.  Blue uses separate empirical age distributions for M-dwarf and FGK hosts, and gray uses their pooled age distribution.  The orange line marks the standardized 5 Gyr median-rotation point estimate, and the dotted line marks $C=0$.}}
\label{fig:historymc}
\end{figure}

\rev{When ages and initial rotations vary among hosts, the median XUV offset is $-0.329^{+0.144}_{-0.143}$ dex.  The median compression is 69.1\%, with a central 95\% range of 55.6\%--82.6\%, and all 2000/2000 realizations give $C>0$.  Drawing both host classes from a pooled age distribution gives a median compression of 71.4\%, a central 95\% range of 58.2\%--84.7\%, and again 2000/2000 realizations with $C>0$ (Figure~\ref{fig:historymc}).  The quoted central ranges are predictive spreads across the stellar histories sampled from the adopted age and rotation priors.  Stellar-history diversity changes the strength of the convergence without changing its direction.}

\subsection{The period-mapping null test}\label{subsec:null}

The period null gives the convergence a direct physical interpretation.  At fixed period, Kepler's law gives $a\propto M_*^{1/3}$, and the activity model assigns a second mass dependence through the integrated stellar XUV energy.  Passing a common period boundary through these two relations can move the M-dwarf and FGK edges closer in XUV even without an explicit atmospheric response.  The null experiment measures how much alignment this coordinate transformation produces for the observed stellar sample.

\begin{figure*}
\centering
\includegraphics[width=0.92\textwidth]{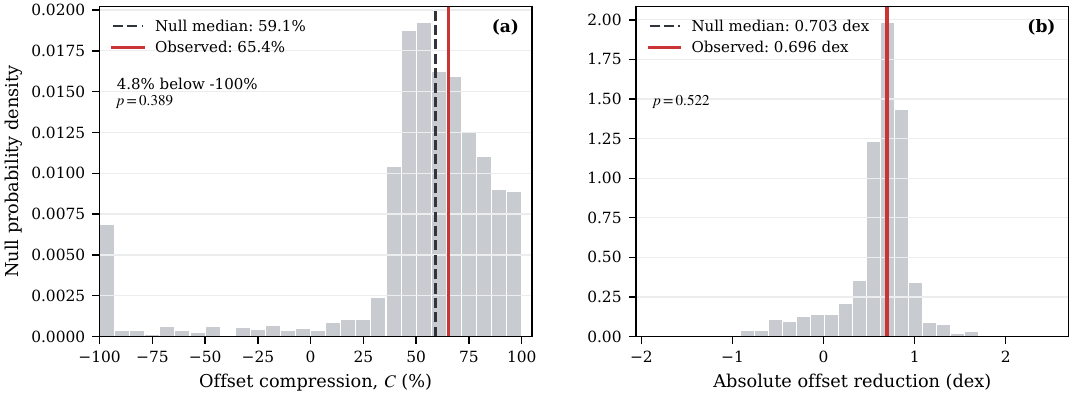}
\caption{Convergence generated by mapping a common period boundary into irradiation coordinates.  The gray distributions contain 2000 realizations with recalculated orbits, irradiation, and weights; red lines mark the observed values.  (a) Fractional compression $C$, with values below $-100\%$ accumulated in the first bin.  (b) Absolute offset reduction $D$, which remains stable when a simulated bolometric offset is small.}
\label{fig:null}
\end{figure*}

In 2000 fully recomputed weighted realizations, the null median is $C=59.1\%$, compared with 65.4\% in the data (Figure~\ref{fig:null}).  A total of 38.9\% of the null realizations are at least as convergent as the observation.  Because the ratio $C$ is sensitive to rare simulations with a nearly zero bolometric offset, we also compare the absolute reduction $D$.  The observed $D=0.696$ dex is close to the null median of 0.703 dex, and 52.2\% of the simulations reach or exceed it.  The measured XUV alignment therefore falls within the range generated by a common period edge and the stellar-mass transformation.

Weighting and multiplanet systems have little effect on this comparison.  In the unweighted null, the observed and median null compressions are 80.3\% and 72.6\% ($p=0.268$), while $D$ gives $p=0.501$.  Retaining one seeded-random planet per host gives 66.4\% and 58.8\% ($p=0.375$), with $p=0.346$ for $D$.  The same physical picture survives both checks: the present population shows XUV convergence, and its magnitude carries no detectable information beyond the common period boundary and the stellar-mass transformation.

\section{Discussion}\label{sec:discussion}

The convergence has a simple physical scale.  Median host properties in the fitted sample are $(M_*,L_*)=(0.50\,\ms,0.041\,L_\odot)$ for M dwarfs and $(0.93\,\ms,0.695\,L_\odot)$ for FGK stars.  At a common period, Keplerian dilution gives an M-to-FGK bolometric-flux ratio of 0.089, or $-1.05$ dex.  The corresponding 5 Gyr median-rotation MORS energies are $5.50\times10^{45}$ and $1.88\times10^{46}$ erg.  Applying the same geometric dilution gives an XUV-fluence ratio of 0.44, or $-0.35$ dex.  These two estimates closely match the fitted offsets.  Low-mass stars emit less total XUV energy, yet their high-energy output is large relative to their bolometric luminosity and persists for longer \citep{Jackson2012,Johnstone2021,Pass2025}.  Replacing current bolometric flux with cumulative XUV removes about 0.7 dex of the apparent spectral-type difference.

\rev{The time dependence is concentrated at young ages and retains a long low-mass tail.  In the median-rotation tracks evaluated at the representative masses above, the first 1 Gyr supplies 55\%--66\% of the 5 Gyr XUV energy and the first 3 Gyr supplies 84\%--88\%.  Solar-mass stars typically leave X-ray saturation early, while late M dwarfs can sustain elevated high-energy emission for billions of years \citep{Tu2015,Johnstone2021,RicheyYowell2023}.  Initial rotation changes the duration of this active phase.  The nine uniform histories therefore span a wide range in compression, and the host-level experiment incorporates this diversity directly.  Its 69.1\% median compression is close to the standardized result, with all 2000/2000 realizations retaining $C>0$.  Stellar-history diversity broadens the allowed amplitude but does not erase the empirical convergence.}

\rev{The host-level experiment remains conditional on MORS, the adopted empirical age priors, and endpoint representation outside the 1--5 Gyr interpolation range.  The M-dwarf prior contains only 25 ages, compared with 538 for FGK hosts.  Its 95\% ranges therefore describe sensitivity within this model and these priors.}

This result extends the lifetime X-ray comparison of \citet{McDonald2019}, which focused on $1.8$--$4\,\re$ sub-Neptunes.  Here the test uses the M-dwarf and FGK desert-opening sample of \citet{Osborn2026}, includes X-rays and EUV, and fits the Kepler--TESS population with one boundary model.  The period-mapping control is the main additional test.  Stellar mass appears in both the MORS energy and $a\propto M_*^{1/3}P^{2/3}$, so a shared period boundary naturally becomes more compact in cumulative XUV.  The observed values of $C$ and $D$ lie near the centers of their null distributions.  The measured convergence does not identify the process that generated the boundary.

Cumulative fluence omits the planet's evolving susceptibility to atmospheric loss.  The magnitude of an energy-limited escape rate can be written
\begin{equation}
 |\dot M_{\rm el}|\simeq
 \frac{\epsilon\,\pi R_p R_{\rm XUV}^{2}F_{\rm XUV}}
 {G M_p K_{\rm tide}},
 \label{eq:mdot}
\end{equation}
where $R_{\rm XUV}$ is the absorbing radius, $\epsilon$ the heating efficiency, and $K_{\rm tide}$ the Roche-potential correction \citep{Erkaev2007,Owen2019}.  Setting $R_{\rm XUV}=R_p$ recovers the familiar $R_p^3$ scaling.  Young planets are larger, so early irradiation acts on a greater cross-section and a shallower binding energy per unit mass.  Core mass, initial envelope fraction, composition, and cooling set how rapidly the radius contracts \citep{LopezFortney2013,OwenWu2013}.  The efficiency also changes with irradiation: highly ionized flows can become recombination-limited, while weakly irradiated flows approach other hydrodynamic regimes \citep{OwenAlvarez2016,Kubyshkina2018}.  A coupled evolution model can thus turn the same $\phix$ into complete stripping, partial loss, or survival.  Its predictions for mass, density, envelope fraction, and age at fixed period provide additional observables that can separate escape from a coordinate mapping.

Orbital history adds another source of structure.  Equation~\ref{eq:fluence} places each planet at its present semimajor axis throughout the integration.  A planet delivered inward after the high-activity epoch would receive a smaller early dose.  High-eccentricity migration and tidal circularization also produce period boundaries that depend on planetary density and arrival time \citep{OwenLai2018,CastroGonzalez2026}.  Joint modeling of escape and migration could explain residual host trends or a broad boundary without invoking one universal fluence.

Survey selection currently limits a population-level physical fit.  Kepler has a measured injection--recovery function, while the TESS entries combine several detection and follow-up pathways.  Separate-survey, unit-weight, and host-deletion tests support the direction of convergence among detected planets.  A selection-complete occurrence measurement requires a homogeneous TESS search or an explicit forward model for each survey.  Additional masses and radii near $2.2$--$7.5\,\re$ would test the binding-energy and density dependences predicted by Equation~\ref{eq:mdot}.

The cleanest discriminant is variation in XUV history at nearly fixed present orbit.  A matched analysis would compare planets with similar $P$, $R_p$, $M_p$, and $M_*$ around stars with independently measured ages, rotation periods, and X-ray or ultraviolet luminosities.  Young clusters, wide coeval binaries, and well-dated field stars provide complementary age anchors; repeating the boundary measurement across age would track the buildup of atmospheric erosion \citep{Gaidos2024}.  The present feasibility screen finds usable age and rotation information for 12 of 33 controlling M-dwarf hosts, with no strict matched pair showing a secure activity contrast.  \rev{The random-history experiment shows that plausible population-level age and rotation diversity does not remove the convergence, but matched systems are still needed to test the XUV histories of individual planets.}  A practical next program should combine a selection-complete planet sample, targeted X-ray and ultraviolet observations, independent age posteriors, and a hierarchical atmosphere-evolution model.  Predictive comparison of a period-only population model with an XUV-plus-atmosphere model would then test whether stellar activity history explains planet radii and masses beyond the geometry already encoded by period.

\section{Conclusions}\label{sec:conclusions}

We tested whether a standardized cumulative XUV dose unifies the empirical lower boundary of the Neptunian desert around M-dwarf and FGK hosts.  The main findings are:

\begin{enumerate}
\item Within the desert-opening radius interval, standardized cumulative XUV removes 65.4\% of the point-estimate host separation seen in present bolometric flux; the host-bootstrap median is 73.4\%.
\item \rev{Assigning separate ages and initial-rotation histories to individual hosts gives a median compression of 69.1\% (95\% range 55.6\%--82.6\%), and all 2000 realizations preserve the convergence.}
\item Period remains the most nearly common empirical coordinate, with XUV intermediate between period and present bolometric flux.
\item A coherently recomputed common-period null naturally reproduces the XUV convergence.  It gives $p=0.389$ for fractional compression and $p=0.522$ for absolute offset reduction; unweighted and one-planet-per-host versions agree.
\item The current detections support XUV as a useful cross-spectral-type coordinate.  Establishing a universal erosion threshold requires homogeneous survey selection, planetary masses, and matched systems with independently constrained activity histories.
\end{enumerate}

\appendix
\restartappendixnumbering
\section{Supporting Diagnostics}\label{app:support}

Table~\ref{tab:objects} summarizes the object-level diagnostic for the direct binned estimator.  TOI-532 has the largest influence on its bolometric edge, whereas TOI-1796.01 has the largest influence on its XUV edge.  Their removal changes the binned compression by less than four percentage points.  In the primary global estimator, leaving out any of 581 hosts gives $C=63.3$--94.2\%.

\begin{deluxetable*}{llcccccc}[t!]
\tablecaption{Highest-impact M-dwarf boundary objects\label{tab:objects}}
\tablehead{
\colhead{Planet} & \colhead{Survey} & \colhead{$R_p/\re$} & \colhead{$P$ (d)} &
\colhead{$M_*/\ms$} & \colhead{$w$} & \colhead{$\phix$ (erg cm$^{-2}$)} &
\colhead{$\Delta C$ (percentage points)}
}
\startdata
TOI-532.01  & TESS & 5.748 & 2.327 & 0.568 & 10.5 & $2.9\times10^{21}$ & $+1.5$ \\
TOI-1796.01 & TESS & 3.522 & 2.644 & 0.380 & 15.2 & $2.1\times10^{21}$ & $-3.5$ \\
TOI-442.01  & TESS & 4.592 & 4.052 & 0.539 & 16.0 & $1.3\times10^{21}$ & $+0.5$ \\
TOI-4526.01 & TESS & 2.597 & 2.260 & 0.298 & 15.4 & $2.7\times10^{21}$ & $-1.9$ \\
TOI-4479.01 & TESS & 2.996 & 1.159 & 0.428 & 8.1  & $6.4\times10^{21}$ & $-1.1$ \\
TOI-674.01  & TESS & 5.225 & 1.977 & 0.398 & 11.9 & $3.1\times10^{21}$ & $-0.7$ \\
\enddata
\tablecomments{Planet names use TESS Object of Interest (TOI) designations. Objects are ranked by the largest leave-one-out change to either the bolometric or XUV boundary. $\Delta C$ is the change relative to the 83.5\% binned baseline; its sign indicates the direction of the change and has no uncertainty meaning.}
\end{deluxetable*}

The median (maximum) weights are 18.24 (55.31), 24.40 (61.39), 11.00 (114.55), and 16.87 (98.46) for Kepler--FGK, Kepler--M, TESS--FGK, and TESS--M, respectively.  Most of the weight is geometric: the detection factor is at its 1.028 floor for the median object in every cell.  This explains why deleting the largest weight is less consequential than deleting an object located directly at an extreme edge.

\begin{figure*}[t!]
\centering
\includegraphics[width=\textwidth]{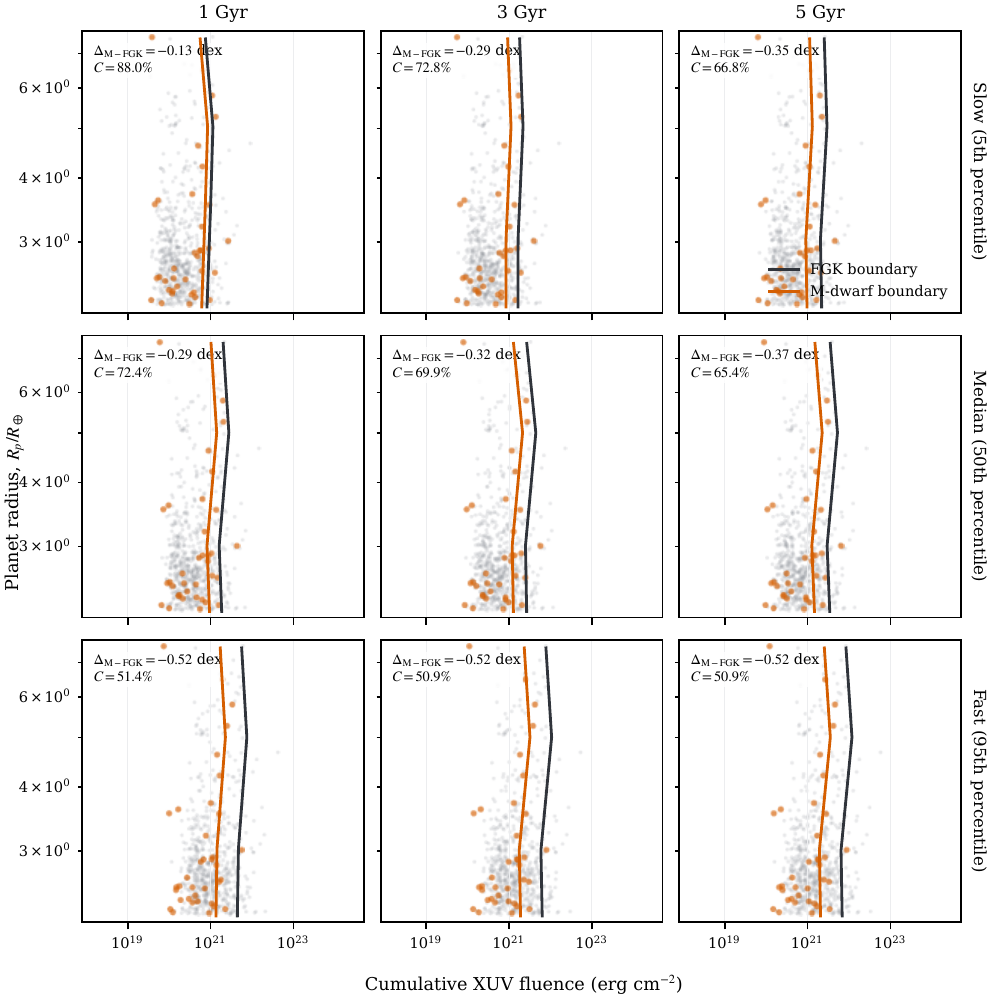}
\caption{\rev{Planet radius versus cumulative XUV fluence for nine uniform stellar-history assumptions.  Columns show integration endpoints of 1, 3, and 5 Gyr; rows show the 5th-, 50th-, and 95th-percentile MORS initial-rotation tracks.  Gray and orange points denote FGK- and M-dwarf-hosted planets in the main $2.24<R_p/\re<7.48$, $P<30$ day sample.  Curves show the $q=0.977$ shared-shape boundary fits.  Each panel lists the M-minus-FGK offset and its compression relative to the present-bolometric-flux offset.  All nine histories yield $C>0$.}}
\label{fig:histories}
\end{figure*}

\begin{center}
\begin{minipage}{0.96\columnwidth}
\centering
\textbf{Table A2. Global-estimator sensitivity checks}\\[3pt]
\footnotesize
\begin{tabular}{lc}
\hline\hline
Variant & $C$ (\%)\\
\hline
Baseline, $2.24$--$7.48\,\re$ & 65.4\\
$2.0$--$8.0\,\re$ & 80.8\\
$2.24$--$5.0\,\re$ & 69.1\\
$q=0.95,0.97,0.99$ & 94.3, 94.8, 59.3\\
Linear radius / one knot / three knots & 60.5, 65.8, 66.5\\
Separate M-dwarf slope & 64.2\\
Unweighted & 80.3\\
Kepler / TESS & 69.2, 93.2\\
Leave one host out & 63.3--94.2\\
Start at 5 / 10 / 20 Myr & 65.3, 65.4, 65.5\\
\hline
\end{tabular}
\end{minipage}
\end{center}

The activity-history feasibility screen contains 37 control planets around 33 M-dwarf hosts.  It combines Kepler rotation catalogs, rotation-based ages \citep{McQuillan2014,Santos2019,Gaidos2024}, identifiers from the TESS Input Catalog (TIC; \citealt{Stassun2019}), and the host-star compilation of \citet{Berger2023}.  \rev{Twenty-five of the 32 M-dwarf hosts in the fitted sample have a reported literature or gyrochronology age in this compilation; the per-object sources are recorded in the archived age-pool table.}  An age passes the precision cut when its mean two-sided relative uncertainty is at most 50\%; a rotation period passes when its relative uncertainty is at most 20\%.  A strict comparison pair must satisfy $|\Delta\log P|\leq0.15$, $|\Delta\log R_p|\leq0.10$, and $|\Delta\log M_*|\leq0.10$.  We require at least 0.30 dex of age or rotation contrast to call the activity histories different.  Twenty-five hosts have a reported age, 21 a rotation period, and 19 both; 12 pass both precision cuts, although several ages are rotation-derived.  Eight strict geometric pairs can be formed; all fall below the activity-history contrast threshold.

Numerical and catalog checks confirm coordinate consistency.  At test masses of 0.155, 0.645, and $1.155\,\ms$, interpolation errors in integrated XUV energy are 0.0066, 0.0054, and $-0.0006$ dex.  \rev{For the age--rotation grid used in the host-level experiment, 16 off-grid MORS calculations give a median absolute interpolation error of 0.0067 dex and a maximum of 0.0161 dex.}  All 44 Kepler--TESS duplicates agree in period to better than $4\times10^{-5}$ fractionally.  Feeding the observed periods through the null-coordinate calculation reproduces catalog detection probabilities to machine precision, weights to a maximum relative error of $5.4\times10^{-15}$, and bolometric fluxes to 0.0033 in maximum relative error.  Period clipping affects 0.44\% of planets in the primary null, and 0.89\% fall below a counterfactual S/N of 7.1.  Ten automated tests cover weighted quantiles, coherent geometry, recovery probability, host resampling, MORS integration, Keplerian mapping, residual permutation, and the age--rotation screen.

\begin{acknowledgments}
\rev{The author thanks the anonymous referee for helpful suggestions.}  This research has made use of the VizieR catalogue access tool, CDS, Strasbourg, France (DOI: 10.26093/cds/vizier).  This paper includes data collected by the Kepler mission.  Funding for the Kepler mission is provided by NASA's Science Mission Directorate.  This paper includes data collected by the TESS mission.  Funding for the TESS mission is provided by NASA's Science Mission Directorate.  This research has made use of the NASA Exoplanet Archive \citep{Christiansen2025}, which is operated by the California Institute of Technology, under contract with the National Aeronautics and Space Administration under the Exoplanet Exploration Program.  Generative-AI tools were used to assist with code implementation and language editing. The scientific analysis, interpretation, and conclusions were independently validated by the author, who takes full responsibility for the content of this work.
\end{acknowledgments}

\facilities{Kepler, TESS}

\software{Matplotlib 3.10.8 \citep{Hunter2007}, MORS \citep{Johnstone2021}, NumPy 2.2.6 \citep{Harris2020}, pandas 2.3.3 \citep{McKinney2010}, SciPy 1.15.3 \citep{Virtanen2020}}

\section*{Data Availability}
The analysis code and derived data products are available on Zenodo at \href{https://doi.org/10.5281/zenodo.23039060}{doi:10.5281/zenodo.23039060}.  The input planet table is available from CDS as catalog J/A+A/709/A23.  MORS is available at \url{https://github.com/FormingWorlds/MORS}.

\bibliographystyle{aasjournalv7}
\bibliography{references}

\end{document}